\documentclass[conference]{IEEEtran}

\usepackage{cite}

\ifCLASSINFOpdf
  \usepackage[pdftex]{graphicx}
\else
  \usepackage[dvips]{graphicx}
\fi
\usepackage[mode=buildnew]{standalone}
\usepackage{pgf, tikz}
\usepackage{pgfplots}
\usepackage{pgfplotstable}
\usepackage{verbatim}
\usetikzlibrary{arrows}
\usetikzlibrary{decorations.text}
\usetikzlibrary{shapes.geometric}
\usetikzlibrary{shapes.arrows}
\usetikzlibrary{arrows}
\usetikzlibrary{shapes}
\usetikzlibrary{positioning}
\usetikzlibrary{shadows}
\usetikzlibrary{patterns}
\pgfplotsset{plot coordinates/math parser=false}
\pgfplotsset{compat=newest}
\pgfplotstableset{use comma,1000 sep=\,}
\pgfkeys{/pgf/number format/.cd,fixed,precision=2}

\usepackage{multirow}

\usepackage[cmex10]{amsmath}
\usepackage{amssymb}

\usepackage{url}

\usepackage[]{fancyhdr} %
\newcommand{\changefont}{\fontsize{9}{9}\selectfont}
\usepackage{url}

\usepackage{color}

\usepackage{import}

\IEEEoverridecommandlockouts
\begin{document}

%
\title{On Aggregation Performance in Privacy Conscious \\Hierarchical Flexibility Coordination Schemes}

\author{\IEEEauthorblockN{Thomas Offergeld\IEEEauthorrefmark{1}\IEEEauthorrefmark{2}, Nils Mattus\IEEEauthorrefmark{1}, Florian Schmidtke\IEEEauthorrefmark{1}, Andreas Ulbig\IEEEauthorrefmark{1}\IEEEauthorrefmark{2}}
\IEEEauthorblockA{\IEEEauthorrefmark{1}Chair of Active Energy Distribution Grids, IAEW, RWTH Aachen University}
\IEEEauthorblockA{\IEEEauthorrefmark{2}Fraunhofer Institute for Applied Information Technology\\
Aachen, Germany\\
\{t.offergeld, f.schmidtke, a.ulbig\}@iaew.rwth-aachen.de}}


%





\maketitle
\thispagestyle{fancy}
\pagestyle{fancy}


\begin{abstract}
  In this paper we introduce a method for performance quantification of flexibility aggregation in flexibility coordination schemes (FCS), with a focus on privacy preserving hierarchical FCS. The quantification is based on two performance metrics: The aggregation error and the aggregation efficiency. We present the simulation framework and the modelling of one complex type of flexibility providing units (FPUs), namely energy storage systems (ESS). ESS cause intertemporal constraints for flexibility coordination that lead to aggregation errors in case flexibility is aggregated from heterogeneous groups of FPUs.

  We identify one parameter responsible for the aggregation error to be the power-to-energy ratio of the ESS. A grouping of FPUs using similarity in their power-to-energy ratios is shown to improve the coordination performance.

  Additionally, we describe the influence of flexibility demand timeseries on the aggregation error, concluding that future assessments of aggregation errors should consider multiple representative demand timeseries, which is a non-trivial task.

  Finally, we discuss the applicability of the developed method to scenarios of larger system size.
\end{abstract}



%
\IEEEpeerreviewmaketitle

\section*{Nomenclature and abbreviations}
\addcontentsline{toc}{section}{Nomenclature}
\noindent\textbf{Elements and Sets}
\begin{IEEEdescription}[\IEEEusemathlabelsep\IEEEsetlabelwidth{$P_{forecast, t, u}$}]
\item[$u, U$] Individual flexibility providing unit (FPU), Set of all FPU
\item[$t, T$] Timestep, Set of all timesteps
\item[$a, A$] Aggregator, Set of all aggregators
\end{IEEEdescription}
\vspace{4pt}
\textbf{Flexibility}
\begin{IEEEdescription}[\IEEEusemathlabelsep\IEEEsetlabelwidth{$P_{forecast, t, u}$}]
\item[$P_{forecast, t, u}$] Forecasted active power
\item[$P_{flex, t, u}$] Flexible active power
\item[$P_{t, u}$] Total active power
\item[$P_{max, u}$] Maximum active power
\item[$P_{req, t, u}$] Requested flexible active power
\item[$P_{IPF, t}$] Interconnection active power flow
\item[$\epsilon_{agg}$] Aggregation error
\item[$\eta_{agg}$] Aggregation efficiency
\end{IEEEdescription}
\vspace{4pt}
\textbf{Energy storage systems}
\begin{IEEEdescription}[\IEEEusemathlabelsep\IEEEsetlabelwidth{$P_{forecast, t, u}$}]
\item[$c_u$] Energy storage system (ESS) capacity
\item[$E_{t, u}$] ESS energy level
\item[$SoC_{t, u}$] ESS state of charge at time $t$ relative to its capacity $c_u$
\item[$P_{chg, t, u}$] Charging power of ESS
\item[$P_{dch, t, u}$] Discharging power of ESS
\item[$X_{chg, t, u}$] ESS charging state ($X\in\{0, 1\}$)
\item[$X_{dch, t, u}$] ESS discharging state ($X\in\{0, 1\}$)
\item[$\eta_{chg, u}$] Charging efficiency
\item[$\eta_{dch, u}$] Discharging efficiency
\item[$PtE_u$] ESS Power-to-Energy ratio
\end{IEEEdescription}
\vspace{4pt}
\textbf{Abbreviations}
\begin{IEEEdescription}[\IEEEusemathlabelsep\IEEEsetlabelwidth{$P_{forecast, t, u}$}]
\item[\textbf{ESS}] Energy storage system
\item[\textbf{FPU}] Flexibility providing unit
\item[\textbf{FCS}] Flexibility coordination scheme
\item[\textbf{FOR}] Feasible operation region
\item[\textbf{IPF}] Interconnection power flow
\end{IEEEdescription}

\section{Introduction}
With the increasing number of distributed energy resources (DER) and energy storage systems (ESS) installed in power systems globally and the simultaneous planned decommissioning of fossil fuelled generation capacity, flexibility of generation and demand is becoming a valuable commodity in the interconnected energy system \cite{hoffrichter_tso-dso-flex, ISEA:MarketReviewStorage}.
Currently, power systems are balanced using different complimentary energy and ancillary service markets where generation and demand is matched to ensure system balance.
With the integration of intermittent renewable generation many systems follow a \textit{renewable-first} approach where the task of balancing the residual demand is left to thermal power plants and storage systems such as pumped hydro and renewables are left uncurtailed when possible.

As a consequence of the ongoing displacement of conventional generation capacity by distributed volatile generation, the procurement of flexibility for system balancing and ancillary services needs to include a large number of small-scale DER.
Besides this market-facing flexibility, DER will be required to provide flexibility for grid operators to counteract congestions in transmission and distribution grids.

The inclusion of a large number of DER, usually connected to sub-transmission and distribution levels, with the required markets and systems used by today's bulk power systems requires novel modelling approaches such as aggregation of resources.
This paper presents a method for the quantification of performance indicators for resource aggregation in hierarchical flexibility coordination schemes (FCS).


These aggregation schemes are oftentimes based on the representation of multiple small flexibilities through a single virtual flexibility. Reasons for aggregation include computational complexity for power system modelling, reductions in data volume for the involved IT systems and privacy for the flexibilities' owners.
However, such an aggregation scheme is only lossless (i.e. it retains all the relevant model information) for homogeneous flexibilities in terms of type, power limits, and for ESS, energy and initial state of charge.
In other cases, the aggregated flexibility can be an over- or underestimation of the available flexibility \cite{kundu}.

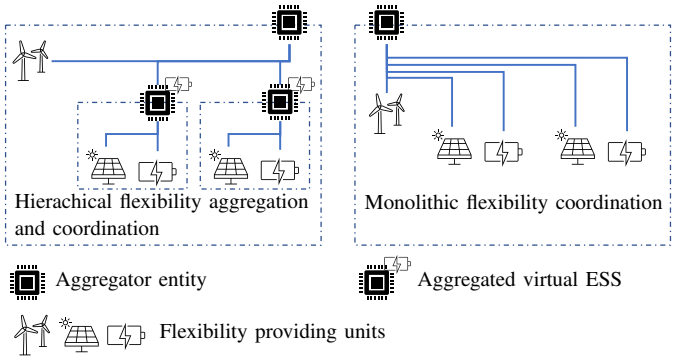
\begin{figure}[t]
    \def\svgwidth{1\columnwidth}
    \fontsize{8}{8}
\begingroup%
  \makeatletter%
  \providecommand\color[2][]{%
    \errmessage{(Inkscape) Color is used for the text in Inkscape, but the package 'color.sty' is not loaded}%
    \renewcommand\color[2][]{}%
  }%
  \providecommand\transparent[1]{%
    \errmessage{(Inkscape) Transparency is used (non-zero) for the text in Inkscape, but the package 'transparent.sty' is not loaded}%
    \renewcommand\transparent[1]{}%
  }%
  \providecommand\rotatebox[2]{#2}%
  \newcommand*\fsize{\dimexpr\f@size pt\relax}%
  \newcommand*\lineheight[1]{\fontsize{\fsize}{#1\fsize}\selectfont}%
  \ifx\svgwidth\undefined%
    \setlength{\unitlength}{1770.43749762bp}%
    \ifx\svgscale\undefined%
      \relax%
    \else%
      \setlength{\unitlength}{\unitlength * \real{\svgscale}}%
    \fi%
  \else%
    \setlength{\unitlength}{\svgwidth}%
  \fi%
  \global\let\svgwidth\undefined%
  \global\let\svgscale\undefined%
  \makeatother%
  \begin{picture}(1,0.52588696)%
    \lineheight{1}%
    \setlength\tabcolsep{0pt}%
    \put(0,0){\includegraphics[width=\unitlength,page=1]{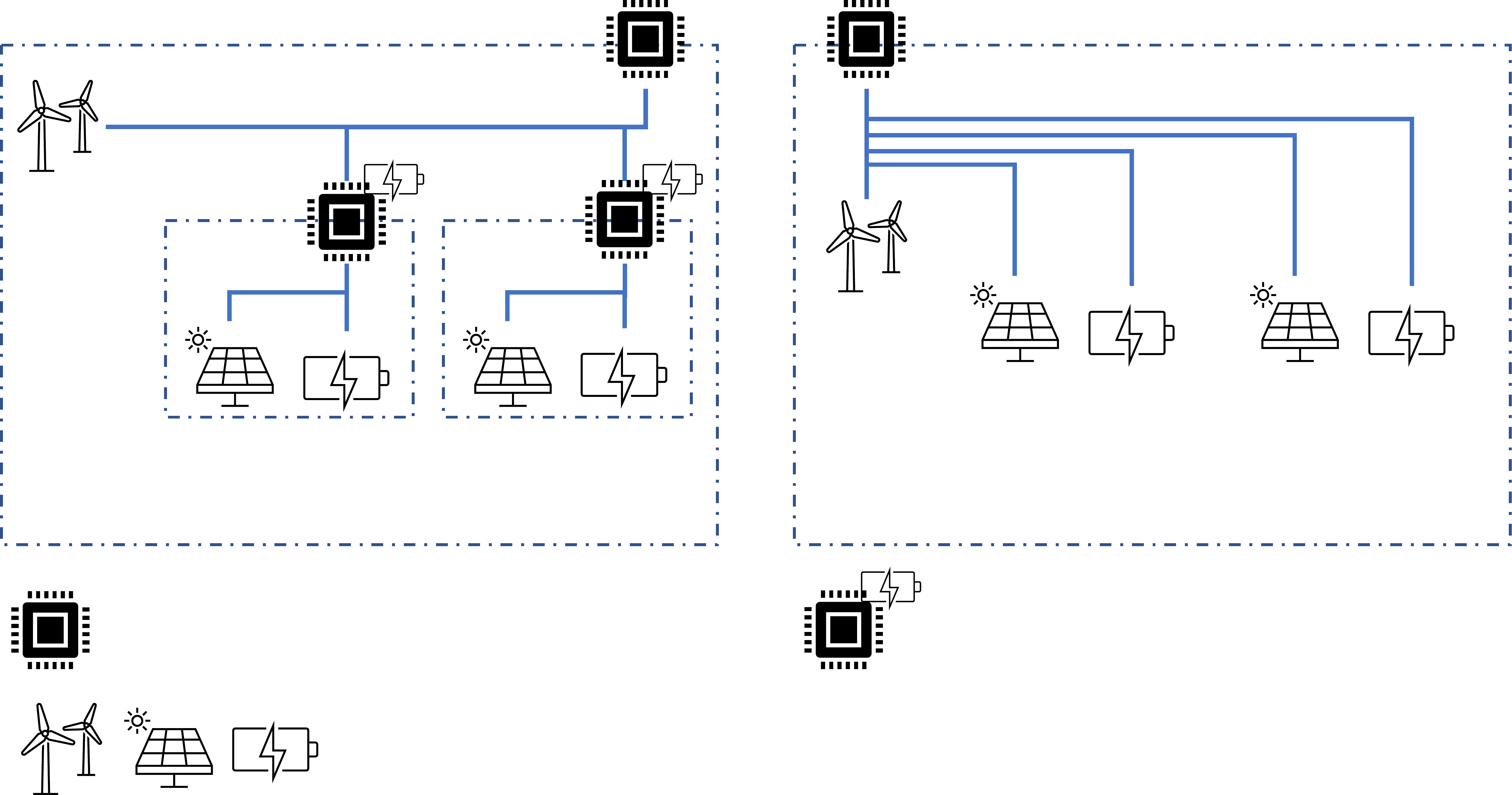}}%
    \put(0.01503728,0.22079899){\color[rgb]{0,0,0}\makebox(0,0)[lt]{\lineheight{1.25}\smash{\begin{tabular}[t]{l}Hierachical flexibility aggregation\\and coordination \end{tabular}}}}%
    \put(0.07597363,0.10366642){\color[rgb]{0,0,0}\makebox(0,0)[lt]{\lineheight{1.25}\smash{\begin{tabular}[t]{l}Aggregator entity\end{tabular}}}}%
    \put(0.61942022,0.10345689){\color[rgb]{0,0,0}\makebox(0,0)[lt]{\lineheight{1.25}\smash{\begin{tabular}[t]{l}Aggregated virtual ESS\end{tabular}}}}%
    \put(0.23257658,0.02655865){\color[rgb]{0,0,0}\makebox(0,0)[lt]{\lineheight{1.25}\smash{\begin{tabular}[t]{l}Flexibility providing units\end{tabular}}}}%
    \put(0.54037344,0.22061363){\color[rgb]{0,0,0}\makebox(0,0)[lt]{\lineheight{1.25}\smash{\begin{tabular}[t]{l}Monolithic flexibility coordination\end{tabular}}}}%
  \end{picture}%
\endgroup%

    \vspace{-18pt}
    \caption{Hierarchical and monolithic flexibility coordination}
    \label{coordination}
    \vspace{-14pt}
\end{figure}

In this paper, we investigate potential sources of aggregation errors by comparing a monolithic FCS with hierarchical FCS with multiple aggregation levels.
Fig. \ref{coordination} shows the logical structure of monolithic and hierarchical FCS in which a possible information flow between different entities is shown by the connecting lines.

Within a monolithic flexibility dispatch, a single power system entity (entitled aggregator) has full access to each flexibilities' characteristics and parameters and retains sole competency to control the flexibility providing units (FPU).
This entity represents a single source of failure for the FCS and is simultaneously tasked with the coordination of potentially millions of individual assets.

On the contrary, in a hierarchical or cascaded FCS, system boundaries are used to abstract information about underlying flexibility potentials, shown by virtual energy system flexibilities in Fig. \ref{coordination}. Note that in this example, the abstracted flexibility potential is represented by a virtual ESS.
This hierarchical approach allows for a more privacy preserving data handling by abstracting information about flexibility potentials at internal system boundaries, for example between plant and grid operators and between different grid operators.

In this paper we provide relevant background information on the challenges and opportunities of flexibility aggregation in future bulk power systems, analyse prior research and deduce the central research questions to be addressed.
Subsequently, we describe a model and method used to investigate performance indicators of aggregation schemes with a focus on coordination of flexibility from ESS.
Finally, we investigate how grouping of individual FPUs within one aggregation level affects the aggregation errors at the expense of model complexity.

\section{Background and State of the Art} \label{chp:background}
\subsection{Flexibility for bulk power systems}
The operation of bulk power systems is built around operational planning processes on different timescales.
For example, within the continental European power system, electricity markets are used to match supply and demand up until the day of delivery (intraday trading) \cite{EC:marketbalancing}.
Besides demand uncertainty, volatile renewable generation introduces uncertainties caused by forecast errors depending mainly on weather effects.
In the past, thermal power plants and pumped hydro storage systems have provided the flexibility required to counteract deviations from the forecasted system balance.
At the same time, the European electricity market is designed to allow equal access to the market for suppliers across its entire domain \cite{EC:marketbalancing}.
However as transmission grids are sometimes incapable of providing the required transmission capacity, European transmission system operators (TSO) are privileged to alter the market result by imposing a re-dispatch of individual generation units to prevent grid congestions.

The current system of this so-called \textit{Redispatch} is used to counteract congestions pre-emptively after electricity market clearing \cite{Amprion:MarketReport2021}.
TSOs determine critical congestions for contingency cases and redispatch the market result in order to guarantee secure grid operation as illustrated in Fig. \ref{redispatch-1}.
Historically, this redispatch only affected large power plants \cite{RD2.0:Girvan}.
However, with the increasing relevance of renewable generation and the congestions caused by large distances between generation and consumption, especially along Germany's north-to-south axis, the system needed to be revised to also include smaller renewable generation units.

Within Germany, grid operators are currently establishing a revised implementation of the \textit{Redispatch} process \cite{RD2.0:BDEW, RD2.0:Muhlpfordt, RD2.0:Girvan}.

\subsection{Redispatch 2.0}

\begin{figure}[t]
    \def\svgwidth{1\columnwidth}
    \fontsize{8}{8}
    \includegraphics[width=\columnwidth]{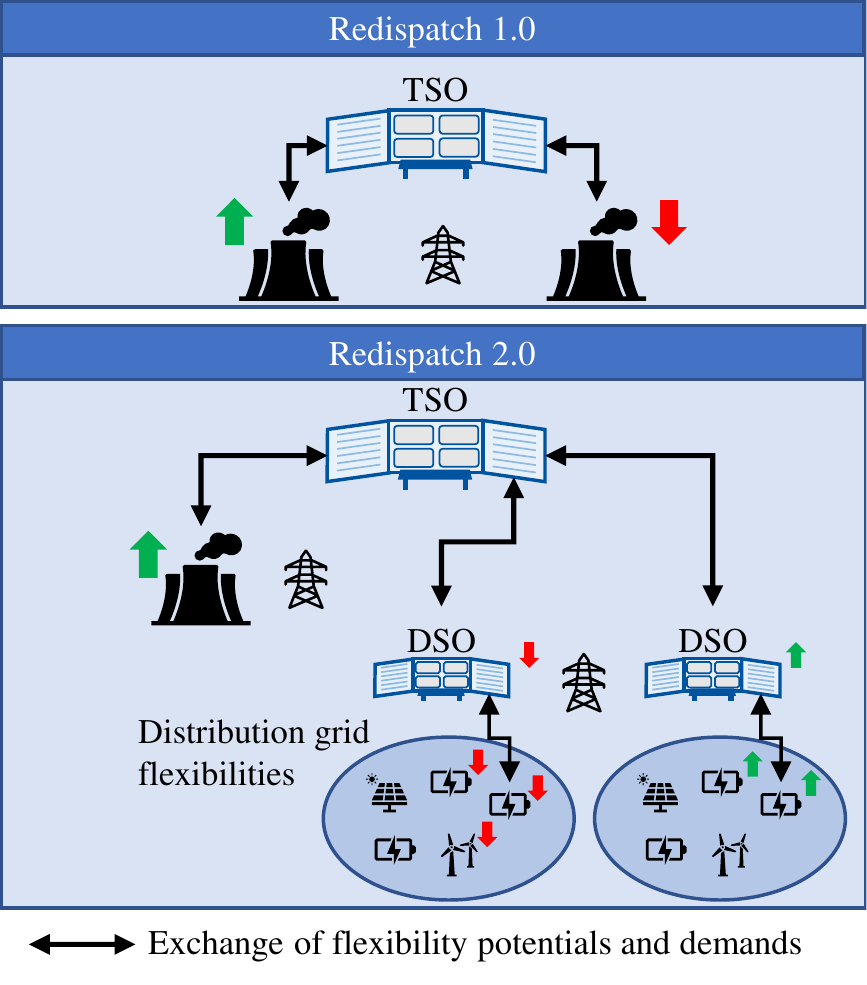}
    \caption{Illustration of conventional Redispatch and Redispatch 2.0 coordinating flexibilities in the bulk power system \cite{RD2.0:BDEW}}
    \label{redispatch-1}
    \vspace*{-12pt}
\end{figure}

With the revised \textit{Redispatch 2.0} process, all generation units and storage systems of 100 kW and more, as well as any DER already controllable by a TSO or DSO, are required to participate \cite{RD2.0:BDEW}.
Other than conventional power plants, these units are predominantly connected to the bulk power systems through distribution grids.
Therefore, a redispatch of these units requires cooperation between TSO and DSO such that a transmission grid motivated redispatch does not cause congestions in the underlying distribution grid.
The process is hierarchical, as plant operators and grid operators on all voltage levels are involved \cite{RD2.0:BDEW}. The revised structure is shown in Fig. \ref{redispatch-1}.
The technical data exchanged between operators comprises active power limits $P_{min}$ and $P_{max}$, forecasted and current injection/consumption $P$, forecasted available positive and negative redispatch potential $P_{red}^+$, $P_{red}^-$, and usable energy for storage systems $E$ and additional data points \cite{RD2.0:BDEW}.

Note that in general there may be multiple DSOs involved in the process between TSO and plant operator.
At each level of the hierarchy, resources can be grouped into one or more resource clusters.
While a single hierarchy level can comprise multiple voltage levels, due to the high fragmentation of the DSO landscape in Germany, with over 800 individual DSO entities, oftentimes each voltage level adds one level of structural hierarchy.

Due to the hierarchical nature of the coordination, limitations of the flexibility can be considered at each level.
On the one hand, flexibility can thus be used to solve congestions in the lower levels of the cascaded system.
On the other hand, congestions caused by flexibility demand from the higher-up levels are avoided by adding constraints to the aggregated flexibility at each level.

\subsection{Definition of Flexibility}

Flexibility is generally understood as a power system asset's ability to alter its power demand or generation set-point in response to an external signal, the flexibility demand.
Each power system resource exhibits limitations of the flexibility it can provide.
Non-flexible resources (NFR) do not provide any flexibility.
Their behaviour is pre-determined and cannot be influenced directly.
An example for NFRs are conventional loads in the power system.
Non-controllable loads oftentimes exhibit voltage-dependent demand reduction that can be exploited using voltage-controlling on-load tap changers \cite{capitanescu_OLTC}.

In the context of this paper, flexibility in the domain of active power defines an FPU's potential to deviate from its unaltered active power injection or consumption behaviour (the \textit{forecasted} behaviour) for a certain amount of time caused by and external signal:
\begin{equation} \label{form:forecast-flex}
    P_{t, u} = P_{forecast, t, u} - P_{flex, t, u}
\end{equation}
where $P_{t, u}$ denotes the mean power consumption or generation between timesteps $t_n$ and $t_{n+1}$.
The interval between two timesteps is defined as $\Delta T$ and the entirety of all timesteps is contained in the set $T$. The forecasted active power $P_{forecast, t, u}$ represents a unit's power demand or generation if no flexibility is used.

Most commonly, the available flexibility is time-dependent \cite{Mayorga:TimeDependentFlexibility}.
The flexibility will furthermore be limited by the capabilities of the FPU (e.g. its minimum and maximum active power $P_{min, u}$ and $P_{max, u}$):
\begin{equation}
    P_{min, u} \leq P_{t, u} \leq P_{max, u}
\end{equation}

For renewable generation in particular, $P_{max, u}$ can be time-variant and limited by the available primary power, e.g. the solar irradiation or wind speed.

This representation can be used to describe flexibility potentials of FPUs where subsequent timesteps are entirely independent of each other.
This can be the case for flexibility from renewable generation units, as long as ramping time-constants are negligible compared to the length of the coordination timesteps \cite{borsche-freq-control-reserve}.

However, ESS in particular introduce time-coupling constraints due to their state of charge and its respective upper and lower limits.
These intertemporal constraints are described in Section \ref{chp3} and are generally non-negligible \cite{Ulbig:Elsevier:OperationalFlexibility}.

\subsection{State of the art}
\subsubsection{Coordination of flexibility}
The coordination of flexibility across different power system actors or voltage levels is a requirement for today's and future power systems.
Distribution grids can provide flexibility at the transmission-distribution interconnect for their overlaying transmission grids.
The same principle can be applied to all kinds of interconnects between parts of the power system.
Other interfaces for the coordination of flexibility are interfaces where different actors are connected (e.g. the TSO-DSO interface, DSO-DSO interfaces, or DSO-prosumer interfaces).

A lot of research has been done on decentralised and hierarchical approaches for flexibility coordination at these interfaces.
In general, they follow a similar understanding of flexibility, that is, the flexibility potential is determined within each entity, communicated to its vertically neighbouring entities, and finally a flexibility request is fulfilled by altering the interconnection power flow (IPF) \cite{Contreras:FeasibleOperatingRange, Contreras:FlexibilityRange, Mayorga:TimeDependentFlexibility, hoffrichter_tso-dso-flex}.

\subsubsection{Aggregation of flexibility potential} \label{SoA:Aggregation}
In recent years, different approaches for the aggregation of flexibility potentials have been developed.
Among these are solutions in which the aggregation result is expressed by the Minkowski sum of individual flexibility polytopes of FPUs \cite{zhao_geometric_aggregation, nrel_union_minkowski_sums, kundu, Ulbig:Elsevier:OperationalFlexibility}.
While Minkowski sums can be computed efficiently for few elements, the computational effort increases exponentially for additional elements and dimensions \cite{Weibel:MinkowskiSums}.
In \cite{kundu}, flexibility potentials are approximated using regular geometric shapes that are transformed using a translation and scaling factor to provide an inner and an outer representation of the flexibility potential before being aggregated.

\begin{figure}[t]
    \vspace{-12pt}
    \def\svgwidth{0.9\columnwidth}
    \fontsize{8}{8}
    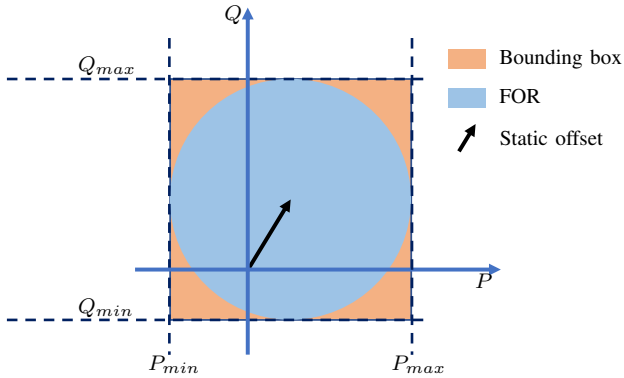
    \caption{Bounding box for two-dimensional sampling}
    \label{FOR-sampling}
    \vspace{-12pt}
\end{figure}

In other work, the feasible operation regions (FOR) for the IPF between active distribution networks (ADN) and overlaying transmissions grids have been determined using random sampling within the two-dimensional P-Q-space \cite{Heleno:RandomSampling:EstimationOfFlexRange, Mayorga:TimeDependentFlexibility}.
The sampled IPF solutions are examined for feasibility.
If there are feasible solutions for the dispatch of the ADN's FPUs, the sample is contained in the FOR.
The convex hull around a large number of feasible samples approximates the flexibility potential of the ADN.
This Monte Carlo sampling is efficient in the two-dimensional P-Q-space if the  independent limit of minimum and maximum P- and Q-flexibility is known.
These limits can be calculated easily from the independent summation of $P$ and $Q$ flexibility potentials of all FPUs.
This concept is shown in Fig. \ref{FOR-sampling} where random samples can be taken from the bounding box.
In this example, the operating range of the FPU is limited by an apparent power limit, resulting in mutually dependent limits of active and reactive power, with an additional inflexible static offset.


In other concepts, approximations of the flexibility potential of an aggregated set of FPUs are represented by a set of linear inequalities.
Using methods of linear optimisation, the extent of the aggregated flexibility potential (the feasible solution space) can be approximated \cite{Contreras:FlexibilityRange, Contreras:FeasibleOperatingRange}.
The time-dependency of the flexibility potential has been identified as a driver for complexity of the aggregation approaches.
This time-dependency exists on different levels:
\begin{itemize}
    \item The time-dependent grid-state and time-dependent flexibility potential of single FPUs \cite{Mayorga:TimeDependentFlexibility, Contreras:FeasibleOperatingRange}
    \item The time-dependency introduced by time-coupling ramping constraints \cite{borsche-freq-control-reserve, Ulbig:Elsevier:OperationalFlexibility}
    \item The time-dependency caused by FPU-internal constraints, such as ESS's minimum and maximum SoC \cite{Ulbig:Elsevier:OperationalFlexibility}
\end{itemize}

\subsection{Research gap and contribution}
While a number of methods for aggregating flexibility potentials exist in literature, the aggregation error introduced by the aggregation of multiple FPUs is not investigated in much detail \cite{Contreras:FlexibilityRange, Mayorga:TimeDependentFlexibility, Contreras:FeasibleOperatingRange, Ulbig:Elsevier:OperationalFlexibility, kundu, hoffrichter_tso-dso-flex, Heleno:RandomSampling:EstimationOfFlexRange}.
While the inaccuracy introduced by abstraction of flexibility potentials into regular geometric shapes is described in \cite{kundu}, the inaccuracy caused by their aggregation is not further investigated.
The aggregation of multiple flexible elements and representation by a virtual, aggregated, element is not lossless in terms of model information for ESS introducing time-coupling constraints.
As we will show later, even the simple aggregation of only four heterogeneous ESS FPUs introduces the risk of overestimating the flexibility potential in the combined virtual FPU.

In order to research mitigation techniques against these inaccuracies, performance metrics for aggregation methods are required.
We will define two different performance metrics, one of which provides a measure of the aggregation error, i.e. the overestimation of flexibility, while the other represents aggregation efficiency, or, how well the physical flexibility can be accessed, compared with a monolithic FCS.

In this paper we will also investigate the following research questions:
\begin{itemize}
    \item Which factors influence the assessment of aggregation performance for hierarchical FCS?
    \item How does the heterogeneity of flexible resources affect the aggregation error and aggregation efficiency?
    \item Which modes or strategies of aggregation can be used to reduce the aggregation error when flexible resources are aggregated into distinct groups?
\end{itemize}

While distribution grid constraints need to be considered for practical applications, we analyse the aggregation of FPUs under the assumption of an uncongested distribution grid. This assumption allows us to focus on the challenges that arise from the aggregation of FPUs. Additionally, in the past, the German transmission grid has had a higher number and severity of congestions as compared to the distribution grids. However, we also acknowledge the relevance of additional distribution grid constraints and discuss their impact on the aggregated flexibility potential.

\section{Modelling Approach} \label{chp3}
In the following we describe the models and the method used to evaluate aggregation characteristics.
The presented method is then used to compare the performance and behaviour of
\begin{itemize}
    \item a monolithic flexibility coordination, and
    \item a hierarchical flexibility coordination scheme with privacy preserving interfaces
\end{itemize}
as shown in Fig. \ref{coordination}.

Besides the simulative evaluation method, component models as well as aggregation and disaggregation tasks are formulated as mixed integer linear optimisation problems using the \textit{gurobi} modelling framework and solver.

\subsection{System architecture for flexibility coordination}
These two approaches to the the flexibility coordination's architecture are defined as follows.

\subsubsection{Monolithic flexibility coordination}
The monolithic or centralised flexibility coordination describes a system in which the full information about all FPUs and their current flexibility potential is available to a centralised authoritative instance.
This instance is able to schedule individual flexibility use from each FPU with the highest degree of confidence.
The monolithic flexibility coordination establishes the benchmark against which a distributed or hierarchical flexibility coordination can be compared.
The monolithic flexibility coordination will always provide a solution at least as good as a system in which the information is not centralised.

However, this approach comes with practical challenges.
The centralised instance presents a single point of failure and requires maintaining communication links with all FPUs.
Additionally, there is no abstraction at system interfaces such as between the asset owner and different system operators.

\subsubsection{Hierarchical flexibility coordination}
The hierarchical FCS is implemented to represent the hierarchical energy system.
The hierarchical coordination enables abstraction layers between different entities and facilitates clustering of FPUs.
This hereby achieved clustering of FPUs into groups of multiple units is privacy preserving as less information about each FPU is collected in a centralised instance and reduces dispatch complexity for the stages higher in the hierarchy by reducing the number of units to be considered.

The hierarchical FCS uses abstraction at its internal boundaries to describe the flexibility potential of the downstream system on the basis of pre-defined building-blocks.
These abstractions are based on the basic FPU models that describe the physical flexibility of the overall system.
The abstractions represent virtual FPUs with properties derived from the downstream system (cf. Fig. \ref{coordination}).
The parametrisation of these virtual FPUs is a fundamental part of the flexibility aggregation.

\begin{figure}[b]
    \vspace{-12pt}
    \def\svgwidth{1\columnwidth}
    \fontsize{8}{8}
\begingroup%
  \makeatletter%
  \providecommand\color[2][]{%
    \errmessage{(Inkscape) Color is used for the text in Inkscape, but the package 'color.sty' is not loaded}%
    \renewcommand\color[2][]{}%
  }%
  \providecommand\transparent[1]{%
    \errmessage{(Inkscape) Transparency is used (non-zero) for the text in Inkscape, but the package 'transparent.sty' is not loaded}%
    \renewcommand\transparent[1]{}%
  }%
  \providecommand\rotatebox[2]{#2}%
  \newcommand*\fsize{\dimexpr\f@size pt\relax}%
  \newcommand*\lineheight[1]{\fontsize{\fsize}{#1\fsize}\selectfont}%
  \ifx\svgwidth\undefined%
    \setlength{\unitlength}{813.50743103bp}%
    \ifx\svgscale\undefined%
      \relax%
    \else%
      \setlength{\unitlength}{\unitlength * \real{\svgscale}}%
    \fi%
  \else%
    \setlength{\unitlength}{\svgwidth}%
  \fi%
  \global\let\svgwidth\undefined%
  \global\let\svgscale\undefined%
  \makeatother%
  \begin{picture}(1,0.61736915)%
    \lineheight{1}%
    \setlength\tabcolsep{0pt}%
    \put(0.01583853,0.20586993){\rotatebox{90}{\makebox(0,0)[lt]{\lineheight{1.25}\smash{\begin{tabular}[t]{l}Aggregation\end{tabular}}}}}%
    \put(0.99534351,0.19881666){\rotatebox{90}{\makebox(0,0)[lt]{\lineheight{1.25}\smash{\begin{tabular}[t]{l}Disaggregation\end{tabular}}}}}%
    \put(0,0){\includegraphics[width=\unitlength,page=1]{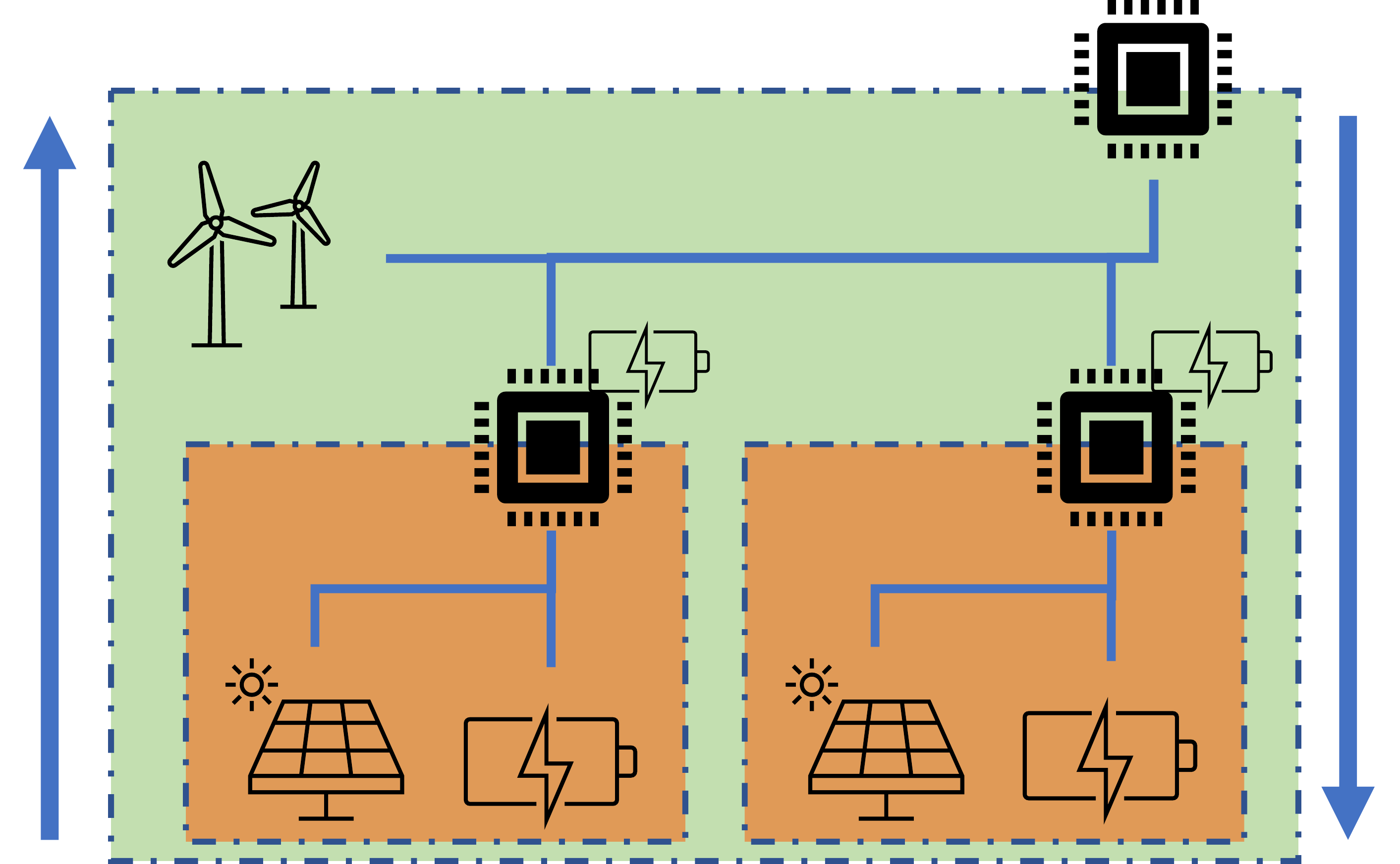}}%
  \end{picture}%
\endgroup%

    \caption{Aggregation and disaggregation in the hierarchical FCS}
    \label{Aggregation_Disaggregation}
\end{figure}

While the determination of the flexibility potential, and the aggregation in the hierarchical FCS, takes place in a bottom-up fashion, the dispatch and disaggregation starts at the highest hierarchy level, with the root entity as shown in Fig. \ref{Aggregation_Disaggregation}.
The aggregation domains are highlighted with two lower aggregation levels shown in orange and the higher aggregation level in green.
Note that the lower aggregation levels are independent of each other and can be computed in a parallel and distributed fashion.
The topmost aggregator, at the same time the only aggregator in the monolithic model, is defined as the root aggregator.

\subsection{Flexibility model}
Flexibilities in the power system can be categorised into time-independent flexibilities and time-dependent flexibilities.

The former describes a set of units where operation between two points in time can be considered independent.
One example for time-independent flexibility is the curtailment of renewable generation units.

The latter describes the more common set of flexibilities, in which an alteration of operating behaviour causes a necessary alteration at a later point in time.
This is true for all demand side management flexibility as the demand still needs to be met but can be shifted in time.
Energy storage systems are also time-dependent as their state of charge depends on the previous state of charge and previous operating point.


Next, we will present the modelling for the energy storage system used to evaluate the aggregation method. The system is modelled using a time-discrete approach where equidistant timesteps $t$ are considered.

The storage system's state of charge $SoC_{t, u}$ is defined relative to the system's capacity $c_u$ and the system's absolute energy content $E_{t, u}$:
\begin{equation}
\begin{gathered}
    0 \leq SoC_{t, u} \leq 1 \quad\forall t \in T \\
    E_{t, u} = SoC_{t, u} \times c_u
\end{gathered}
\end{equation}
The system can either be charged, discharged or be idle.
This is enforced using binary variables $X_{chg}$ and $X_{dch}$ taking the value of $1$ if the charging or discharging process is active, respectively.
These equations hold for each timestep $t \in T$. The charging and discharging states are mutually exclusive.
\begin{equation} \label{form:charge-discharge}
\begin{gathered}
    0 \leq P_{chg, t, u} \leq X_{chg, t, u} \times P_{max, u} \\
    0 \leq P_{dch, t, u} \leq X_{dch, t, u} \times P_{max, u} \\
    X_{chg, t, u} + X_{dch, t, u} \leq 1
\end{gathered}
\end{equation}
Finally, the continuity equation ensures that the state of charge is dependent on the previous state of charge
\begin{equation}
\begin{aligned}
    SoC_{t_i, u} ={} & SoC_{t_{i-1}, u} + \\
                   & \left(P_{chg, t, u} \times \eta_{chg, u} - \frac{P_{dch, t, u}}{\eta_{dch, u}}\right) \times \frac{\Delta T}{c_u}
\end{aligned}
\end{equation}
where $\eta_{chg}$ and $\eta_{dch}$ are the charging and discharging efficiency of the system and $\Delta T$ is the time between timesteps $t_{i-1}$ and $t_i$.

The power injection or demand of the ESS is described by the difference of charging and discharging power, at least one of which is zero due to (\ref{form:charge-discharge})
\begin{equation}
    P_{t, u} = P_{chg, t, u} - P_{dch, t, u},
\end{equation}
and links $P_{chg, t, u}$ and $P_{dch, t, u}$ to the flexibility of this system given by (\ref{form:forecast-flex}).

\subsection{Flexibility Aggregation} \label{chp:model:aggregation}
The FPUs grouped by an aggregator are aggregated into one or more virtual FPUs.
For the FPUs, modelled as storage systems, the active power and capacity are aggregated by summation, whereas the aggregation of the state of charge involves weight coefficients equal to each ESS capacity:
\begin{equation}
\begin{gathered}
    P_{a, t} = \sum_{u \in U} P_{t, u} \\
    c_{a} = \sum_{u \in U} c_u \\
    SoC_{a, t} = \frac{\sum_{u \in U} SoC_{u, t} \times c_u}{\sum_{u \in U} c_u}
\end{gathered}
\end{equation}

To a higher-level aggregator, an aggregator behaves identically to an individual FPU. Aggregators are typed using the FPU types, for example an aggregator containing only ESS-FPUs would be typed as a virtual ESS. Therefore, aggregators and FPUs can be combined by aggregation.

The aggregation of FPUs can be based upon different criteria. One such criterion is geographic or electrical proximity of FPUs.
However, further criteria can be applied within a set of FPUs selected for aggregation based on proximity information.
Given a set of multiple storage systems with non-identical parameters ($P_{max}$, $c$ and initial $SoC$), the aggregation modes are:
\begin{itemize}
    \item Aggregate all FPUs into one virtual FPU
    \item Aggregate similar FPUs into individual virtual FPUs
    \item Aggregate dissimilar FPUs into individual virtual FPUs
\end{itemize}

The second and third option are defined as homogeneous and heterogeneous aggregation modes respectively.
In order to investigate the impact of these aggregation strategies on the aggregation error, both will be compared in Section \ref{chp:results}.

\subsection{Disaggregation}
The disaggregation describes the dispatch of flexibilities, in the hierarchical and monolithic flexibility coordination, according to its objective function.

The objective for the disaggregation is to minimise the quadratic difference between the requested flexibility $P_{req, t}$ and the delivered flexibility $P_{flex, t}$:
\begin{equation}
\label{Disagg:Obj}
    \min \sum_{t \in T} (P_{req, t} - P_{flex, t})^2
\end{equation}

For the monolithic flexibility coordination, a single step of disaggregation is sufficient to determine all FPUs' operating points, while within the hierarchical flexibility coordination, a disaggregation is calculated for each aggregator. In that case, the requested flexibility $P_{req, t}$ is a result of the disaggregation of the next-higher aggregator.

\subsection{Aggregation metrics}
In order to assess the performance of different approaches of aggregation, we define two metrics:
\begin{itemize}
    \item The aggregation error $\epsilon_{agg}$
    \item The relative aggregation efficiency $\eta_{agg}$
\end{itemize}

Both metrics provide an indication of how closely the aggregation result models the exact flexibility potential of the aggregated system.

\subsubsection{Aggregation error}
The aggregation error describes the loss of information of the aggregation, evident by a mismatch of the flexibility requested and delivered.
\begin{equation}
\label{agg_error}
    \epsilon_{agg} = \dfrac{\sum_{t \in T} (P_{req, t} - P_{flex, t})^2}{\sum_{t \in T} (P_{req, t})^2}
\end{equation}

The aggregation error is proportional to the objective value for the disaggregation step described in (\ref{Disagg:Obj}) and can be calculated for each aggregator, including the root aggregator.
An aggregation error larger than zero identifies a case where the process of aggregation was not lossless and the aggregator determined a flexibility dispatch that was infeasible due to constraints unknown to the aggregator. During disaggregation, the infeasible dispatch is followed as closely as possible by minimising the quadratic difference of the objective function (\ref{Disagg:Obj}).

\subsubsection{Aggregation efficiency}
The aggregation efficiency is a metric of how efficiently flexibility can be used in the hierarchical FCS compared to the monolithic FCS.
An efficiency of less than $100\%$ denotes that the monolithic FCS was able to make better use of the system's total flexibility than the hierarchical system.
In mathematical terms, the efficiency is the ratio of flexibility used by the hierarchical system, compared to the monolithic system (if no flexibility is provided in the monolithic system, the efficiency is not defined):
\begin{equation}
    \eta_{agg} = \dfrac{\sum_{t \in T} |P_{flex, hierarchic, t}|} {\sum_{t \in T} |P_{flex, monolithic, t}|}
\end{equation}

\subsection{Simulation framework}
The simulation framework is built upon the \textit{pandapower} library for power system simulation that provides the data model used to determine the aggregation groups \cite{pandapower.2018}.
Although distribution grid congestions are not considered, a practical flexibility hierarchy can be derived from the grid structure. For example, in a flexibility coordination ranging from individual household energy systems to a medium voltage grid, aggregators are defined at two system boundaries
\begin{itemize}
    \item The household connection point
    \item The MV/LV substation
\end{itemize}

\begin{figure}[t]
    \def\svgwidth{1\columnwidth}
    \fontsize{8}{8}
    \includegraphics[width=\columnwidth]{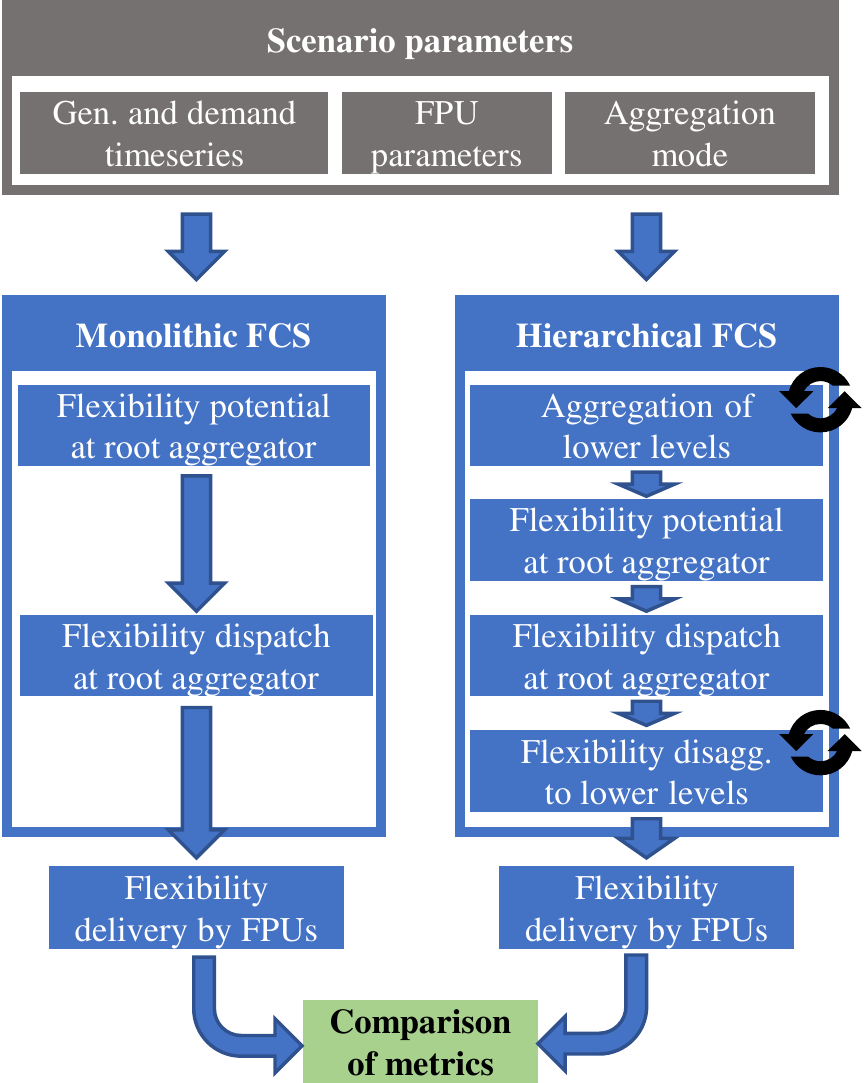}
    \caption{The simulation procedure including hierarchical and monolithic FCS}
    \label{sim-structure}
    \vspace{-12pt}
\end{figure}

The framework structure is shown in Fig. \ref{sim-structure}.
Each simulation scenario is defined by the aggregation mode, the FPU and aggregation parameters, defining the characteristics of the FPUs and the hierarchical structure, and finally the dynamic generation and demand timeseries that cause fluctuating flexibility potential and demand.
If elements are not congestion-free, these additional limitations can be imposed as part of the aggregation. In addition to the modelling described up to this point, such constraints can limit the power flow between the cluster of resources and the aggregator:
\begin{equation}
    P_{t, u} \leq P_{constraint, t}
\end{equation}


For both the monolithic and hierarchical FCS, the root aggregator provides the only interconnection to the external system.

The hierarchical aggregation and later disaggregation is only used in the hierarchical FCS.
Both the monolithic and hierarchical FCS calculate the flexibility potentials and the dispatch and invoke the flexibility delivery accordingly.
The aggregation and disaggregation is repeated for each aggregator instance within the hierarchical FCS while the monolithic FCS uses a single aggregation and disaggregation of all FPUs under the root aggregator.

\section{Exemplary Results} \label{chp:results}
In the following section we will present results from simulating the flexibility dispatch of the hierarchical FCS and the monolithic FCS.

\subsection{Impact of aggregation mode}
\begin{figure}[t]
    \def\svgwidth{1\columnwidth}
    \input{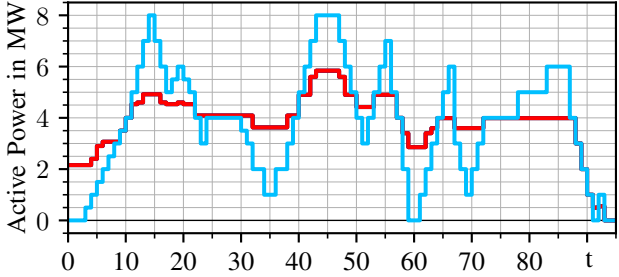}
    \vspace*{-48pt}
    \caption{Homogeneous aggregation}
    \label{homogeneous}
    \vspace*{-12pt}
\end{figure}

To illustrate the challenges caused by the independent aggregation of FPU parameters, we aggregate a total of four ESS with non-identical parameters using the aggregation modes defined in Section \ref{chp:model:aggregation}, illustrated by Fig. \ref{heterogeneous}, \ref{homogeneous} and \ref{identical_pte}. The diagrams show the IPF realised by the root aggregator, which is the sum of the static, non-flexible active power and the delivered active power flexibility $P_{flex, t, u}$.
In order to clearly demonstrate the effect, we have chosen two ESS with large power-to-energy (PtE) ratios ($3.25$), while the others have low PtE ratios ($0.4375$). A PtE equal to one, defined by
\begin{equation}
\label{eq:pte}
    PtE_u = \dfrac{P_{max, u}}{c_u}
\end{equation}
describes an ESS that can be fully charged, or discharged, within one hour.
The parameters of the ESS are shown in Table \ref{tab:ess_param}.
The horizon of flexibility coordination is sampled into 96 timesteps, spanning 24 hours with $\Delta T = 0.25h$.

Deviations between the hierarchical optimised IPF ($P_{IPF_{hier}}$) and the monolithic IPF ($P_{IPF_{mon}}$) suggest a suboptimal efficiency of the hierarchical flexibility coordination.
The monolithic flexibility coordination is always at least as optimal as the hierarchical flexibility coordination due to the additional detailed information of the FPUs available.

A difference between the planned hierarchical IPF ($P_{IPF_{hier, plan}}$) and the realised hierarchical IPF ($P_{IPF_{hier}}$) is a sign of aggregation errors.
The planned IPF can not be realised due to constraints at lower levels of the hierarchy unknown to the root aggregator.

\begin{table}[b]
    \vspace{-12pt}
    \centering
    \begin{tabular}{|c|c|c|c|c|}
    \cline{3-5}
    \multicolumn{2}{c|}{} & \multicolumn{3}{c|}{Scenario} \\
    \cline{3-5}
    \multicolumn{2}{c|}{} & (a) & (b) & (c) \\
    \cline{3-5}
    \multicolumn{2}{c|}{} & \multicolumn{3}{c|}{\textbf{Aggregation mode}} \\
    \hline
    Group & FPU & Het. & Hom. & {Identical PtE} \\
    \hline
    \hline
    \multirow{4}{2em}{1} & \multirow{2}{1em}{1} & 1.3 MW & 1.3 MW & 0.4 MW \\
    & & 0.4 MWh & 0.4 MWh & 0.4 MWh \\
    \cline{2-5}
    & \multirow{2}{1em}{2} & 0.7 MW & 1.3 MW & 1.6 MW \\
    &&  1.6 MWh & 0.4 MWh & 1.6 MWh \\
    \hline
    \multirow{4}{2em}{2} & \multirow{2}{1em}{3} & 1.3 MW & 0.7 MW & 0.4 MW \\
    & & 0.4 MWh & 1.6 MWh & 0.4 MWh \\
    \cline{2-5}
    & \multirow{2}{1em}{4} & 0.7 MW & 0.7 MW & 1.6 MW \\
    &&  1.6 MWh & 1.6 MWh & 1.6 MWh \\
    \hline
    \hline
    \multicolumn{2}{|c|}{Total power} & 4 MW & 4 MW & 4 MW \\
    \hline
    \multicolumn{2}{|c|}{Total capacity} & 4 MWh & 4 MWh & 4 MWh \\
    \hline
    \end{tabular}
    \vspace{6pt}
    \caption{ESS parameters}
    \vspace{-12pt}
    \label{tab:ess_param}
\end{table}

\subsubsection{Heterogeneous aggregation} \label{chp:results:heterogeneous}
\begin{figure}[t]
    \def\svgwidth{1\columnwidth}
    \input{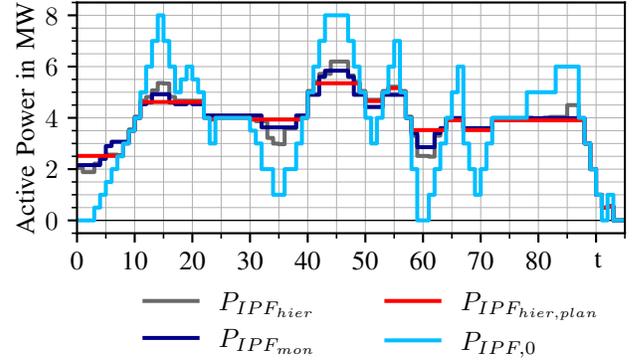}
    \vspace*{-6pt}
    \caption{Heterogeneous aggregation}
    \label{heterogeneous}
    \vspace*{-12pt}
\end{figure}
In Fig. \ref{heterogeneous} we observe the original IPF at the root aggregator without any flexibility use in \textbf{light blue} colour as $P_{IPF, 0}$.
The optimised flexibility dispatch with the objective of achieving an optimised IPF in the monolithic coordination scheme is shown in \textbf{dark blue} ($P_{IPF_{mon}}$).
This reference case describes the optimal flexibility dispatch of a monolithic coordinator with all information about the FPUs.
The optimisation of the root aggregator's IPF follows the objective function
\begin{equation}
\label{Obj:FourSto}
    \min\sum_{t=t_0}^{t_{max}} P_t^2
\end{equation}
which realises a minimisation of the quadratic sum of the interconnection active power flow.
Note that the specific objective function is used to generate the synthetic flexibility demand required to prompt a flexibility dispatch from the FCS, however, differing objective functions are  possible.

Aggregating the parameters of the FPUs heterogeneously (scenario (a) in Table \ref{tab:ess_param}), that is, one ESS with high PtE ratio is combined with one of low PtE ratio, we observe that the flexibility potential of the combined virtual FPU is overestimated.
The \textbf{red} graph indicates the planned IPF of the root aggregator ($P_{IPF_{hier, plan}}$) which can be compared against the globally optimal dispatch of the monolithic coordinator (dark blue).
The planned flexibility is disaggregated to the two aggregators each containing two FPUs with heterogeneous parameters.
Other than the root aggregator, these aggregators have all information about the underlying flexibilities that are needed to determine that the flexibility requested by the root aggregator is infeasible. The flexibility is therefore provided using the best-effort disaggregation that minimises the quadratic deviation between the requested and provided flexibility defined in (\ref{Disagg:Obj}).
Finally, the \textbf{grey} line shows the IPF with the best-effort flexibility dispatch in the hierarchical FCS.

The resulting aggregation error in the heterogeneous aggregation is $0.055$.
The aggregation efficiency, comparing the effect of the flexibility realised in the hierarchical and monolithic FCS, is $87.5\%$.

For the heterogeneous aggregation and subsequent disaggregation we conclude:
\begin{itemize}
    \item The aggregation error of the monolithic FCS is zero. The IPF planned by the monolithic FCS can be fulfilled.
    \item The root aggregator of the hierarchical FCS is lacking information about the distribution of power and energy among the FPUs as this information is abstracted at the first aggregator level. The root aggregator overestimates the available flexibility.
    \item The overestimation of available flexibility by the root aggregator causes an inefficient flexibility dispatch. The hierarchical FCS uses less flexibility overall and achieves an objective value that is worse than the monolithic FCS.
\end{itemize}

\subsubsection{Homogeneous aggregation}
The behaviour of a homogeneous aggregation of FPUs (scenario (b) in Table \ref{tab:ess_param}) on the error and efficiency is shown in Fig. \ref{homogeneous}.

The reference dispatch generated by the monolithic flexibility coordination is identical to the dispatch in Fig. \ref{heterogeneous} as the aggregation mode remains identical for the monolithic FCS.
In case of homogeneous aggregation of flexibility resources, the resulting flexibility dispatch matches the reference dispatch exactly for the hierarchical FCS as well. Thus the coordination efficiency of the homogeneous aggregation is $100\%$.
Clearly, the identical parameters of the paired ESS in the homogeneous aggregation mode cause a beneficial behaviour when aggregated.

Additionally, the realised flexibility for the hierarchical FCS in the homogeneous case is equal to the planned flexibility, hence the aggregation error as defined by (\ref{agg_error}) is zero. Note that, corresponding to $\eta_{agg} = 100\%$ and $\epsilon_{agg} = 0$, $P_{IPF_{mon}}$ and $P_{IPF_{hier}}$ are both equal to $P_{IPF_{hier, plan}}$ and are therefore not visible in Fig. \ref{homogeneous}.

\begin{table}[b]
    \vspace{-12pt}
    \centering
    \begin{tabular}{c|c|c|c|}
    \cline{2-4}
    {} & \multicolumn{3}{c|}{Scenario} \\
    \cline{2-4}
    {} & (a) & (b) & (c) \\
    \cline{2-4}
    {} & \multicolumn{3}{c|}{\textbf{Aggregation mode}} \\
    \cline{2-4}
    {} & Het. & Hom. & {Identical PtE} \\
    \cline{2-4}
    \hline
    \multicolumn{1}{|c|}{Aggregation error $\epsilon_{agg}$} & 0.055 & 0 & 0 \\
    \hline
    \multicolumn{1}{|c|}{Aggregation efficiency $\eta_{agg}$} & 87.5\% & 100\% & 100\% \\
    \hline
    \end{tabular}
    \vspace{6pt}
    \caption{Aggregation error and efficiency}
    \vspace{-12pt}
    \label{tab:ess_results}
\end{table}

\subsubsection{Heterogeneous aggregation with identical PtE}
\begin{figure}[t]
    \def\svgwidth{1\columnwidth}
    \input{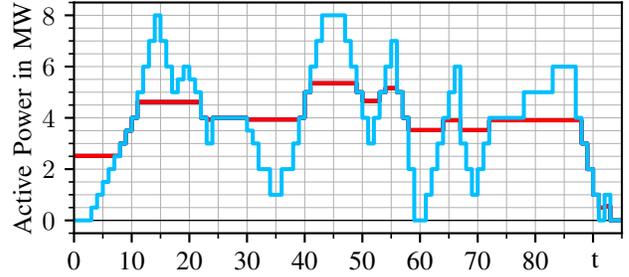}
    \vspace*{-48pt}
    \caption{Heterogeneous aggregation with identical PtE ratio}
    \label{identical_pte}
    \vspace*{-12pt}
\end{figure}
Based on a hypothesis that the aggregation error is caused by dissimilar ratios of power to energy capacity in the ESS, we investigate one more purposefully constructed scenario. We create an heterogeneous aggregation mode of ESS with identical PtE ratios but different power limits and capacities (scenario (c) in Table \ref{tab:ess_param}). While all ESS exhibit identical PtE, the ESS are still grouped into two groups.
The result illustrated in Fig. \ref{identical_pte} shows a scenario in which the PtE ratio is one for all ESS, however the overall flexibility in terms of power and energy remains unchanged.

When comparing the monolithic flexibility dispatch, we observe a higher use of flexibility compared to scenarios (a) and (b).
This behaviour can be attributed to the fact that the altered distribution of power and energy among the FPUs is more beneficial for the given flexibility demand. The ESS with low PtE used in (a) and (b) required several hours to charge or discharge entirely.

We observe that the heterogeneous aggregation of ESS FPUs with identical PtE ratios is also lossless as seen in Fig. \ref{identical_pte} ($\eta_{agg}=100\%$, $\epsilon_{agg}=0$).
Additionally, the overall objective value improves slightly due to the altered distribution of power and energy among the ESS.

In light of the research questions identified in Section \ref{chp:background} we have identified that the ratio of power and energy capacity of ESS is a potential source of aggregation errors. At the same time, a heterogeneous ESS power or energy alone does not appear to increase aggregation errors.
It can be concluded that a potential method to reduce the impact of aggregation errors can be a clustering based on similarity in terms of power-to-energy ratio.

\subsection{Influence of varying storage parameters}
Having identified a varying PtE as key driver for aggregation errors, we investigate the impact a variation of the PtE ratio has on the aggregation error.

\begin{figure}[t]
    \def\svgwidth{1\columnwidth}
    \hspace{-12pt}
    \input{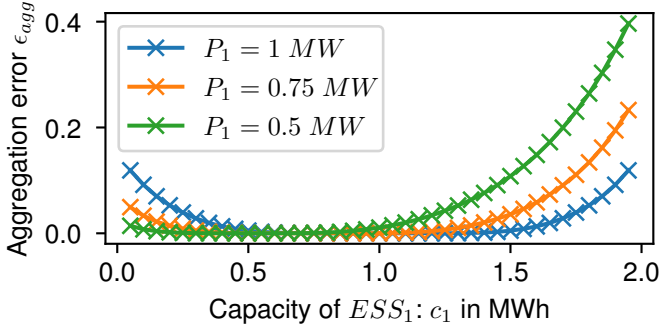}
    \vspace{-24pt}
    \caption{Change of aggregation error $\epsilon$ with changing $c_1$}
    \label{variation_pte}
    \vspace{-12pt}
\end{figure}

Building upon the scenario investigated in Section \ref{chp:results:heterogeneous} the distribution of storage capacity within each group of ESS is varied. The impact of varying capacity distribution among the storage systems within one aggregation group can be seen in Fig. \ref{variation_pte}. The diagram shows the aggregation error $\epsilon_{agg}$ for a changing distribution of the total capacity of one group ($2\ MWh$) between two ESS.

The capacity of one storage system ($ESS_1$) is varied from $0.05\ MWh$ to $1.95\ MWh$. For $c_1 = 0\ MWh$ and $c_1 = 2\ MWh$, no aggregation takes place, therefore the aggregation error is zero. The capacity of the second storage system is defined as
\begin{equation} \label{c_2}
    c_2 = 2\ MWh - c_1
\end{equation}

Additionally, the maximum and minimum power of each storage system is varied. As with the capacity, the total power of the group is fixed at $2\ MW$ while the power of the first ESS is varied. The maximum power of the second system is calculated as
\begin{equation}
    P_{max, 2} = P_2 = 2\ MW - P_1
\end{equation}

Shown is the change of aggregation error $\epsilon$ for $P_1 = \{0.5, 0.75, 1\}\ MW$.

For $P_1 = P_2 = 1\ MW$ we observe symmetry in results due to (\ref{c_2}) with $\epsilon_{c_1 = 0.5\ MWh} = \epsilon_{c_1 = 1.5\ MWh}$, however, the symmetry does not hold if the power distribution is also uneven. In addition, the aggregation error exhibits a low gradient for scenarios where less than 75\% of capacity is allocated to one system for $P_1 = P_2$. Changing the balance of the storage power causes this gradient dead-band to shift.

\subsection{Influence of flexibility demand}
\begin{figure}[t]
    \def\svgwidth{1\columnwidth}
    \hspace{-12pt}
    \input{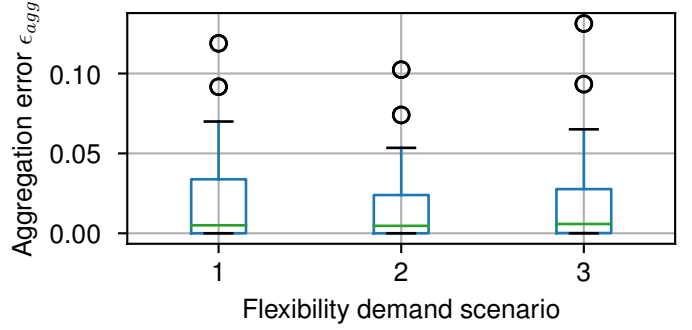}
    \vspace{-24pt}
    \caption{Impact of changing flexibility demand timeseries}
    \label{flex_demand_impact}
    \vspace{-12pt}
\end{figure}

As described, the flexibility demand disaggregated by the root aggregator is generated synthetically. In Fig. \ref{flex_demand_impact} we see the impact of three different exemplary but not representative flexibility demand timeseries for a flexibility coordination between storage systems with equal power but unequal capacity. It is apparent that the aggregation error is affected by the flexibility demand that is disaggregated. An assessment of aggregation errors should therefore be extended to be based on multiple representative flexibility demand timeseries going forward.

However, determining representative flexibility demand timeseries is a non-trivial challenge that warrants future investigation. Potential approaches include a geometric sampling of the FOR, which will be a space of $\mathbb{R}^n$ for scenarios with $n$ timesteps when considering active power only, methods for which are developed in \cite{Kiatsupaibul:Hit-and-Run,Chen:MCMC-sampling}.

\subsection{Discussion on computational effort of larger simulations}
Going beyond the examples shown in this paper requires more computational effort as the number of aggregators and FPUs increases. While the examples shown up until here were computationally inexpensive and simulated in simulation times below two seconds, simulations of larger scenarios are prolonged.

For a scenario based on a synthetic benchmark case for power distribution grids from the simbench project containing 28 FPUs, 5 aggregators and a two-day optimisation horizon, resulting in 192 individual timesteps, the simulation of the hierarchical FCS takes less than 60 seconds to compute on office equipment \cite{simbench_mdpi}.

Implementing hierarchical FCS can be realised in a highly distributed way, allowing the parallel computation of all aggregations and disaggregations on each aggregation level.
With the advent of edge computing capabilities close to DERs, such a decentralisation of computation is possible.

However, for the structured benchmarking of different FCS against each other, centralised simulation provides a simplified way for data handling.

\section{Conclusion and future work}
\subsection{Conclusion}
We have introduced a method to evaluate aggregation schemes for hierarchical flexibility coordination against a reference FCS with perfect system insight.
Based on a mixed integer linear programming approach to model the operational flexibility of energy storage systems we develop a model for the aggregation of flexible resources.
Clustering and aggregation of flexible resources will be a requirement of future bulk power system operation processes to manage computational complexity and increase data privacy of flexibility providers.
We introduce two metrics for aggregation performance: the aggregation efficiency $\eta_{agg}$ and aggregation error $\epsilon_{agg}$.

It is observed that aggregation performance is affected by FPUs' characteristics and parameters, however an aggregation error is caused only by heterogeneous power-to-energy ratios of ESS. Differences in power or energy capacity alone do not appear to cause aggregation errors in the hierarchical FCS.
A strategy to lower the aggregation error is to group FPUs with similar PtE ratios together whenever possible.

Additionally, the assessment of aggregation errors is dependent on the flexibility demand that is disaggregated by the system. For future investigations, a range of flexibility demand timeseries should be considered.

We show that a privacy conscious coordination of flexibility is possible by aggregating multiple flexible resources into virtual resources. However, careless aggregation can result in aggregation errors and inefficiencies. The hierarchical coordination reduces the insight that central coordinators need to maintain into the individual FPUs.

\subsection{Future work}
Building upon the method developed in this contribution, further aggregation techniques can be developed and evaluated.

In future work we will investigate the impact of flexibility demand on the aggregation error.
In this contribution, we utilised representative flexibility demand timeseries from generation and consumption timeseries.
In order to benchmark an FCS we will look at synthetic flexibility demands that sample the feasible space of flexibility potential more uniformly.
While the efficiency of Monte Carlo samplings deteriorates for a high-dimensional solution space, other approaches such as hit-and-run sampling may provide a tractable method for sampling the FOR \cite{Kiatsupaibul:Hit-and-Run}.

Besides the aggregation of energy resources, grid constraints can be added to both the hierarchical and monolithic flexibility coordination scheme.
While the combination of flexibility potentials and grid constraints has been demonstrated in numerous publications, such as \cite{Mayorga:TimeDependentFlexibility}, the impact on aggregation errors warrants further investigation.

The models for flexibility providing units can be extended further by additional degrees of freedom, such as reactive power provision.
Furthermore, energy storage systems also exhibit nonlinear behaviour that can be included.


\bibliographystyle{IEEEtran}
\bibliography{References/refs}

\begin{thebibliography}{10}
\providecommand{\url}[1]{#1}
\csname url@samestyle\endcsname
\providecommand{\newblock}{\relax}
\providecommand{\bibinfo}[2]{#2}
\providecommand{\BIBentrySTDinterwordspacing}{\spaceskip=0pt\relax}
\providecommand{\BIBentryALTinterwordstretchfactor}{4}
\providecommand{\BIBentryALTinterwordspacing}{\spaceskip=\fontdimen2\font plus
\BIBentryALTinterwordstretchfactor\fontdimen3\font minus
  \fontdimen4\font\relax}
\providecommand{\BIBforeignlanguage}[2]{{%
\expandafter\ifx\csname l@#1\endcsname\relax
\typeout{** WARNING: IEEEtran.bst: No hyphenation pattern has been}%
\typeout{** loaded for the language `#1'. Using the pattern for}%
\typeout{** the default language instead.}%
\else
\language=\csname l@#1\endcsname
\fi
#2}}
\providecommand{\BIBdecl}{\relax}
\BIBdecl

\bibitem{hoffrichter_tso-dso-flex}
A.~Hoffrichter, T.~Offergeld, A.~Blank, and T.~Kulms, ``Simulation of
  transmission grid operation incorporating flexibility at distribution
  level,'' in \emph{16th International Conference on the European Energy
  Market}, 2019.

\bibitem{ISEA:MarketReviewStorage}
\BIBentryALTinterwordspacing
J.~Figgener, C.~Hecht, D.~Haberschusz, J.~Bors, K.~G. Spreuer, K.-P. Kairies,
  P.~Stenzel, and D.~U. Sauer, ``The development of battery storage systems in
  germany: A market review (status 2022),'' 2022. [Online]. Available:
  \url{https://arxiv.org/abs/2203.06762}
\BIBentrySTDinterwordspacing

\bibitem{kundu}
S.~Kundu, K.~Kalsi, and S.~Backhaus, ``Approximating flexibility in distributed
  energy resources: A geometric approach,'' in \emph{2018 Power Systems
  Computation Conference (PSCC)}, 2018.

\bibitem{EC:marketbalancing}
{European Commission}, ``Guideline on electricity balancing,'' 2017.

\bibitem{Amprion:MarketReport2021}
{Amprion GmbH}, ``Market report 2021,'' 2021.

\bibitem{RD2.0:Girvan}
P.~Girvan, A.~Stolte, and M.~Mann, ``Conceptualising different approaches to
  the new redispatch procedure in {Germany} with special regard to the
  distribution of competences among the grid operators,'' in \emph{PESS 2020;
  IEEE Power and Energy Student Summit}, 2020.

\bibitem{RD2.0:BDEW}
\BIBentryALTinterwordspacing
{BDEW Bundesverband der Energie- und Wasserwirtschaft e.V.},
  ``{BDEW-Branchenlösung Redispatch 2.0},'' 2020. [Online]. Available:
  \url{https://www.bdew.de/service/anwendungshilfen/bdew-branchenloesung-redispatch-20/}
\BIBentrySTDinterwordspacing

\bibitem{RD2.0:Muhlpfordt}
T.~Mühlpfordt, X.~Dai, A.~Engelmann, and V.~Hagenmeyer, ``Distributed power
  flow and distributed optimization—formulation, solution, and open source
  implementation,'' \emph{Sustainable Energy, Grids and Networks}, vol.~26,
  2021.

\bibitem{capitanescu_OLTC}
F.~Capitanescu, ``{AC OPF-Based Methodology for Exploiting Flexibility
  Provision at TSO/DSO Interface via OLTC-Controlled Demand Reduction},'' in
  \emph{2018 Power Systems Computation Conference (PSCC)}, 2018.

\bibitem{Mayorga:TimeDependentFlexibility}
D.~Mayorga~Gonzalez, J.~Hachenberger, J.~Hinker, F.~Rewald, U.~Häger,
  C.~Rehtanz, and J.~Myrzik, ``Determination of the time-dependent flexibility
  of active distribution networks to control their {TSO-DSO} interconnection
  power flow,'' in \emph{2018 Power Systems Computation Conference (PSCC)},
  2018.

\bibitem{borsche-freq-control-reserve}
T.~Borsche, A.~Ulbig, and G.~Andersson, ``A new frequency control reserve
  framework based on energy-constrained units,'' in \emph{2014 Power Systems
  Computation Conference}, 2014.

\bibitem{Ulbig:Elsevier:OperationalFlexibility}
A.~Ulbig and G.~Andersson, ``Analyzing operational flexibility of electric
  power systems,'' \emph{International Journal of Electrical Power \& Energy
  Systems}, vol.~72, 2015, the Special Issue for 18th Power Systems Computation
  Conference.

\bibitem{Contreras:FeasibleOperatingRange}
D.~Contreras and K.~Rudion, ``Computing the feasible operating region of active
  distribution networks: Comparison and validation of random sampling and
  optimal power flow based methods,'' \emph{IET Generation, Transmission \&
  Distribution}, vol.~15, 2021.

\bibitem{Contreras:FlexibilityRange}
D.~A. Contreras and K.~Rudion, ``Improved assessment of the flexibility range
  of distribution grids using linear optimization,'' in \emph{2018 Power
  Systems Computation Conference (PSCC)}, 2018.

\bibitem{zhao_geometric_aggregation}
L.~Zhao, W.~Zhang, H.~Hao, and K.~Kalsi, ``A geometric approach to aggregate
  flexibility modeling of thermostatically controlled loads,'' \emph{IEEE
  Transactions on Power Systems}, vol.~32, no.~6, 2017.

\bibitem{nrel_union_minkowski_sums}
M.~S. Nazir, I.~A. Hiskens, A.~Bernstein, and E.~Dall'Anese, ``Inner
  approximation of minkowski sums: A union-based approach and applications to
  aggregated energy resources,'' in \emph{2018 IEEE Conference on Decision and
  Control (CDC)}, 2018.

\bibitem{Weibel:MinkowskiSums}
C.~Weibel, ``Minkowski sums of polytopes combinatorics and computation,'' 2007.

\bibitem{Heleno:RandomSampling:EstimationOfFlexRange}
M.~Heleno, R.~Soares, J.~Sumaili, R.~Bessa, L.~Seca, and M.~A. Matos,
  ``Estimation of the flexibility range in the transmission-distribution
  boundary,'' in \emph{2015 IEEE Eindhoven PowerTech}, 2015.

\bibitem{pandapower.2018}
L.~Thurner, A.~Scheidler, F.~Sch{\"a}fer, J.~Menke, J.~Dollichon, F.~Meier,
  S.~Meinecke, and M.~Braun, ``pandapower — an open-source python tool for
  convenient modeling, analysis, and optimization of electric power systems,''
  \emph{IEEE Transactions on Power Systems}, vol.~33, no.~6, 2018.

\bibitem{Kiatsupaibul:Hit-and-Run}
S.~Kiatsupaibul, R.~L. Smith, and Z.~B. Zabinsky, ``An analysis of a variation
  of hit-and-run for uniform sampling from general regions,'' \emph{ACM Trans.
  Model. Comput. Simul.}, vol.~21, no.~3, feb 2011.

\bibitem{Chen:MCMC-sampling}
Y.~Chen, R.~Dwivedi, M.~J. Wainwright, and B.~Yu, ``Fast {MCMC} sampling
  algorithms on polytopes,'' \emph{Journal of Machine Learning Research},
  vol.~19, no.~55, pp. 1--86, 2018.

\bibitem{simbench_mdpi}
S.~Meinecke, D.~Sarajlić, S.~R. Drauz, A.~Klettke, L.-P. Lauven, C.~Rehtanz,
  A.~Moser, and M.~Braun, ``Simbench—a benchmark dataset of electric power
  systems to compare innovative solutions based on power flow analysis,''
  \emph{Energies}, vol.~13, no.~12, 2020.

\end{thebibliography}

\end{document}